\documentclass[english,aps,pra,groupedaddress,nofootinbib,reprint,onecolumn,notitlepage]{revtex4-1}
\usepackage{color}
\usepackage{babel}
\usepackage{array}
\usepackage{hyperref}
\usepackage{float}
\usepackage{amsthm}
\usepackage{amsmath}
\usepackage{amssymb}
\usepackage{graphicx}
\usepackage{bm}

\newenvironment{proofof}[1]{\begin{trivlist}\item[]{\flushleft\it
Proof of~#1. }}
{\qed\end{trivlist}}

\makeatletter

\providecommand{\tabularnewline}{\\}
\floatstyle{ruled}
\newfloat{algorithm}{tbp}{loa}
\providecommand{\algorithmname}{Algorithm}
\floatname{algorithm}{\protect\algorithmname}

\newfloat{algorithmclass}{tbp}{loa}
\providecommand{\algorithmnameb}{Algorithm Class}
\floatname{algorithmclass}{\protect\algorithmnameb}

\@ifundefined{textcolor}{}
{%
 \definecolor{BLACK}{gray}{0}
 \definecolor{WHITE}{gray}{1}
 \definecolor{RED}{rgb}{1,0,0}
 \definecolor{GREEN}{rgb}{0,1,0}
 \definecolor{BLUE}{rgb}{0,0,1}
 \definecolor{CYAN}{cmyk}{1,0,0,0}
 \definecolor{MAGENTA}{cmyk}{0,1,0,0}
 \definecolor{YELLOW}{cmyk}{0,0,1,0}
 }
\theoremstyle{plain}

  \theoremstyle{definition}

\@ifundefined{definecolor}
 {\usepackage{color}}{}

\usepackage{amsfonts}\usepackage{amsthm}\usepackage{bbm}

\@ifundefined{definecolor}{\usepackage{color}}{}
\usepackage{pifont}\newcommand{\cmark}{\ding{51}}%
\newcommand{\xmark}{\ding{55}}%

\usepackage{subfigure}

\newcommand{\op}[2]{|#1\rangle\!\langle#2|}

\newcommand{\ket}[1]{\left|#1\right\rangle}
\newcommand{\bra}[1]{\left\langle #1 \right|}

\newcommand{\arearegion}{\mathcal{A}}
\newcommand{\area}{|\mathcal{A}|}

\newcommand{\barn}{\overline{n}}

\newcommand{\naivename}{(2)}
\newcommand{\measj}[1]{\mathcal{M}(V_{\mathcal{M},#1})}
\newcommand{\measz}{\mathcal{M}(V_{\mathcal{M}})}

\newcommand{\Tr}{\mathrm{Tr}}

\newcommand{\registerword}{mode}

\newcommand{\Prsim}{\Pr_{{\rm sim}, \measz}}
\newcommand{\Prtruncquant}{\Pr_{{\rm trunc-quant},\measz}}
\newcommand{\Prquant}{\Pr_{{\rm quant},\measz}}
\newcommand{\Prsimsamp}{\Pr_{\text{{\rm sim}},\rho}}

\newtheorem{definition}{Definition}

\newtheorem{theorem}{Theorem}

\graphicspath{{./}{./figures/}}

\providecommand{\definitionname}{Definition}
\providecommand{\theoremname}{Theorem}

\makeatother

  \providecommand{\definitionname}{Definition}
\providecommand{\theoremname}{Theorem}

\begin{document}
\global\long\def\ket#1{\left| #1 \right\rangle }

\global\long\def\bra#1{\left\langle #1 \right|}

\global\long\def\braket#1#2{\left\langle #1 | #2 \right\rangle }

\global\long\def\ketbra#1#2{|#1\rangle\!\langle#2|}

\global\long\def\braopket#1#2#3{\bra{#1}#2\ket{#3}}

\global\long\def\Tr{\text{Tr}}

\global\long\def\Pr{\text{Pr} }

\title{Efficient simulation scheme for a class of quantum optics experiments with non-negative Wigner representation}


\author{Victor Veitch$^{1,2}$}

\author{Nathan Wiebe$^{1,3}$}

\author{Christopher Ferrie$^{1,2}$}

\author{Joseph Emerson$^{1,2}$}

\affiliation{$^{1}$Institute for Quantum Computing, University of Waterloo, Waterloo,
Ontario, Canada, N2L 3G1}

\affiliation{$^{2}$Department of Applied Mathematics, University of Waterloo,
Waterloo, Ontario, Canada, N2L 3G1}

\affiliation{$^{3}$Department of Combinatorics \& Opt., University of Waterloo,
Waterloo, Ontario, Canada, N2L 3G1}


\date{\today}
\begin{abstract}
We provide a scheme for efficient simulation of a broad class of quantum optics experiments. Our efficient simulation  extends the
continuous variable Gottesman-Knill theorem to a lage class of non-Gaussian mixed states, thereby identifying
that these non-Gaussian states are not an enabling  resource
for exponential quantum speed-up. 
Our resuls also provide an operationally motivated
interpretation of negativity as non-classicality.  We apply our scheme to the case of noisy single-photon-added-thermal-states to show that this class admits states with positive Wigner function but negative $P$-function that are not useful resource states for quantum computation. 
\end{abstract}
\maketitle

\section{Introduction}

There have been a variety of approaches to the problem of characterizing what is non-classical about quantum theory.  There are many important signatures of quantum theory, but with the rise of quantum information, the exponential speedups in quantum algorithms over the best known classical algorithms have increasingly become an important signature of quantum theory.  This point is especially poignant in light of recent work by Aaronson and Arkhipov wherein a simple non--universal linear optical system is shown to be able to perform computational tasks believed to be hard for classical computers~\cite{Aaronson2010Computational}.  The extent to which computational speedups and the boundaries between computational complexity classes are reflected in more traditional measures of non--classicality remains, however, an open question.  In continuous variable quantum theory, and quantum optics in particular, the most frequently considered notions of quantumness are phrased in terms of the so-called quasi-probability distributions, such as the Wigner function and the (Glauber-Sudarshan) $P$-function.  There is a strong tradition in physics of considering negativity of the quasi-probability function as an indicator of non--classicality of a quantum state \cite{Mandel1986NonClassical,Paz1993Reduction,Bell2004Speakable,Kalev2009Inadequacy}.  It is therefore natural to suspect that negativity is intimately linked to computational speedups in both discrete and continuous quantum information processing.


Continuous variable quantum information theory provides a potentially
powerful alternative to the usual discrete formalism and many of the
seminal results in discrete variable quantum computation have analogs
in the continuous variable setting. 
Perhaps the most important example
is the {}``continuous variable Gottesman-Knill theorem'', which
states that a computation restricted to the subset of quantum theory
containing only Gaussian states and operations is classically efficiently
simulatable \cite{Bartlett2002Efficient,Bartlett2002Efficient2}.
More concretely, unitary Gaussian quantum information is defined to
be the following set of operators (see, for example, \cite{Weedbrook2011Gaussian}):
$n$ mode Gaussian input state; quadratic Hamiltonians; and, measurements
with (or without) post-selection onto Gaussian states. Bartlett \emph{et
al.} \cite{Bartlett2002Efficient,Bartlett2002Efficient2} have shown
explicitly that there exists a classical algorithm that reproduces the
output probabilities of the measurement results that executes in time
that scales polynomially with the number of modes. This shows 
that some non-Gaussian resources are necessary to obtain an exponential speed-up with quantum optical experiments, but leaves open the question of whether they are sufficient.   

Recently, Veitch \emph{et al.}~\cite{Veitch2012Negative} have shown that
 a discrete analog of the Wigner function ~\cite{Gross2006Hudsons} can be used to define a necessary
condition for a mixed quantum state to enable an exponential speed-up through quantum computation. 
Their model considers the use of Clifford operations on \emph{qudits}
(the case of qubits is not covered by their proof) and  measurements of stabilizer states and finds, somewhat surprisingly, that there exist a class of \emph{bound universal states} outside of the convex set of stabilizer states that can still be efficiently simulated and therefore do not serve as a resource for exponential speed-up with quantum computation. 
There is  a tight mathematical correspondence between the discrete and continuous Wigner representations, the Clifford/stabilizer model for qudits \cite{Gottesman1997Stabilizer}  and  the Gaussian model for quantum optics considered
by Bartlett \emph{et al.}~\cite{Bartlett2002Efficient,Bartlett2002Efficient2}.
It is therefore natural to ask whether the restriction to Gaussian states in the model of Bartlett
\emph{et al.} can be relaxed to allow more general class of initial states that have non-negative
Wigner representation while still permitting an efficient classical
simulation. 

This work affirms an answer in the positive by showing that a large
class of quantum states with positive Wigner representation exists outside the convex hull of the
$n$-mode Gaussian states that can be efficiently simulated using a classical computer, given
restrictions to quadratic Hamiltonians and Gaussian measurement. This shows, in particular, that linear optical
quantum devices are essentially no more computationally powerful than classical computers under such restrictions.  We show this by providing
an explicit classical simulation algorithm that can be used to simulate
sampling the output probability distributions of the evolved initial states. As a practical application we apply our results to determine a threshold on the computational power of single-photon-added-thermal states (SPATS)  \cite{Agarwal1992Nonclassical,Zavatta2004QuantumtoClassical,Parigi2007Probing} for variable efficiencies.   In this sense, our work serves as both a conceptual and practical generalization of the continuous variable Gottesman--Knill theorem to a broader class of input states.

This paper is outlined as follows. We begin with a brief review of
the Wigner function formalism and Gaussian quantum mechanics in section~\ref{sec:wigner}. We then provide
our simulation protocol for states with positive Wigner representation
 in section~\ref{sec:sim}.  In section~\ref{sec:SPATS}, we discuss positivity
of the Wigner function as a necessary condition for quantum computation.
We illustrate the bound state region via the recently studied class
of limited-efficiency SPATS
and show that quantum efficiencies of 50\% are a necessary threshold
for computational speed-up. Finally, section~\ref{sec:conclusion} contains our conclusion and further discussion about our findings.

\section{Review of Wigner Functions}\label{sec:wigner}

Wigner functions provide a natural quantum analog of the classical
phase space distribution of a dynamical system. 
We provide
below a brief review of the properties of Wigner functions. For simplicity,
we focus our attention on Wigner functions for a single particle (or
equivalently a single mode) in one dimension. The generalization to
higher dimensions and more particles is straightforward~\cite{Wigner1932On}.

The Wigner representation of a state $\rho$ is defined to be~\cite{Wigner1932On}
\begin{equation}
W_{\rho}(q,p)=(2\pi)^{-1}\int_{-\infty}^{\infty}\bra{q-y/2}\rho\ket{q+y/2}e^{ipy/\hbar}\mathrm{d}y,
\end{equation}
 where $\ket q$ is a position eigenstate. 
 The Wigner function is both positive and negative in general. However,
it otherwise has many of the same properties as a classical probability density on
phase space. For these reasons, the Wigner function is often referred
to as a \emph{quasi-probability} function. Intuitively, if we could
find a bona fide joint probability distribution of non-commuting observables,
then there would be no difference between quantum and classical theories.
It is not surprising, then, that \emph{negativity} is necessary in
all possible quasi-probability representations of a quantum state
\cite{Ferrie2010Necessity}.

The time--evolution of the Wigner function for a Hamiltonian of the
form $H=p^{2}/2m+V(q)$ is given by~\cite{Wigner1932On,balbook}
\begin{equation}
\frac{\partial W_{\rho}(q,p)}{\partial t} =\{H,W_\rho\} + \sum_{\ell=1}^\infty\frac{1}{(2\ell+1)!}\left(-\frac i2\right)^{2\ell}\frac{\partial^{2\ell+1}V(q)}{\partial q^{2\ell+1}}\frac{\partial^{2\ell+1}W_{\rho}(q,p)}{\partial {p}^{2\ell+1}},
\end{equation}
where $\{\cdot,\cdot\}$ is the Poisson bracket, which governs classical Hamiltonian equations of motions. 
This result is important because it states that the
time--evolution of $W_{\rho}(q,p)$ is given by Liouville's equation,
plus a quantum correction.  The quantum correction is zero for the
case of the quadratic Hamiltonian:
\begin{equation}\label{eq:Wpevol}
\frac{\partial W_{\rho}(q,p)}{\partial t}  = \{H,W_\rho\}.
\end{equation}
Hence the evolution equation
agrees \emph{precisely} with the classical predictions. This observation
will be vital for our simulation algorithm because the Hamiltonians
permitted by linear optics are quadratic (Harmonic oscillators), which (along
with our non--negativity assumption) implies that we will be able
to simulate the evolution of $W_{\rho}(q,p)$ using an ensemble of
classical trajectories.

This discussion above implies that a Wigner function that is initially
classical, meaning that it is non--negative and hence interpretable
as a probability density function (pdf), will remain classical under the
action of a quadratic Hamiltonian. In this context  is therefore useful to determine the conditions under which a Wigner function is
non--negative as this gives a practically relevant boundary between quantum and classical
states. Hudson's theorem \cite{Hudson1974When} was the first attempt
to characterize the positive Wigner functions and it was later generalized
to the following \cite{SotoEguibar1983When}. Let $\psi$ be a pure quantum state of $n$ oscillators
(modes). Then its Wigner function is positive if and only if 
\begin{equation}
\psi(\vec{Q})=e^{-\frac{1}{2}(\vec{Q}\cdot A\vec{Q}+B\cdot\vec{Q}+c)},
\end{equation}
 where $A$ is an $n\times n$ Hermitian matrix, $B$ is an $n$-dimensional
complex vector and $c$ is a normalization constant. In quantum
optics terminology, these are either \emph{coherent states} or
\emph{squeezed states}.  That is, plugging these states into the definition of the
Wigner function yields multivariate Gaussian distributions in phase
space. Convex combinations of these states (incoherent mixtures of them)
also have positive Wigner function since the mapping is linear. Early
on, these were incorrectly conjectured to be the only such mixed states
with positive Wigner function. The question of mixed states was given
a full treatment in reference \cite{Srinivas1975Some} and later in
\cite{Brocker1995Mixed}. Both references independently found that
a theorem in classical probability attributed to Bochner \cite{Bochner1933Monotone}
and generalization thereof can be used to characterize both the valid
Wigner functions and the subset of positive ones. What was shown is
that there exist a large class of states with positive Wigner function
that are not convex combinations of Gaussian states. So far, these
states have received little attention. In Section~\ref{sec:SPATS}, we show  that such states are more than a mathematical
curiosity; they arise naturally in quantum optics.

Gaussian measurements are also easily modeled in the Wigner
representation. Recall that for non-negative states the Wigner function picture allows us to represent the system as a probability density over underlying physical states in phase space, $\bm{u}_{f}$.
Gaussian measurements in this picture are also modeled as probability densities
for outcomes $\bm{k}$, conditioned on the value of the underlying physical state $\bm{u}_{f}$.
Specifically, consider the case of measurement $\mathcal{M}$ of a Gaussian
state $G$ with covariance matrix $V_\mathcal{M}$. The POVM representation of this
measurement is \cite{Leonhardt1998Measuring}
\begin{equation}
\measz=\{\ketbra{G(\bm{k},V_{\mathcal{M}}}{G(\bm{k},V_{\mathcal{M}})}\ :\ \bm{k}\in\mathbb{R}^{2n}\},\label{eq:guass_meas_povm}
\end{equation}
which selects a Gaussian state with mean $\bm{k}$ and covariance
$V_\mathcal{M}$ from all possible Gaussian states with mean $k$ and covariance $V_\mathcal{M}$. In the
Wigner function picture the representation of this measurement is,
\begin{equation}
M_{\measz}(\bm{k}|\bm{u}_f)=\frac{1}{\mathcal{X}}\exp\left(-(\bm{k}-\bm{u}_f)^{T}V_\mathcal{M}^{-1}(\bm{k}-\bm{u}_f)\right),\label{eq:gauss_meas_wig_rep}
\end{equation}
where $\mathcal{X}$ is the normalization constant. We introduce the notation in~\eqref{eq:gauss_meas_wig_rep} to emphasize the difference
between the representations of measurements and states. The interpretation
of this equation is that if the system is actually at the point $\bm{u}_f$
the effect of a measurement will be to produce an outcome
$\bm{k}$ according to the probability density $M_{\measz}(\bm{k}|\bm{u}_f)$.
Using this equation and the law of total probability we can find the
probability density of measurement outcomes $\bm{k}$ for
the measurement $\measz$ on a quantum state with Wigner representation
$W_{\rho}(\bm{k})$: 
\begin{equation}
p(\bm{k}|\measz,\rho)=\int_{\bm{u}}W_\rho(\bm{u})M_{\measz}(\bm{k}|\bm{u})d\bm{u}.
\end{equation}
Of course, this agrees with the probability assigned by the Born rule
\begin{equation}
p(\bm{k}|\measz,\rho)=\Tr(\ketbra{G(\bm{k},V_{\mathcal{M}}}{G(\bm{k},V_{\mathcal{M}})}\rho).
\end{equation}

The simulation algorithm that we propose uses none of the special properties of Gaussian measurements other than
the fact that they have a positive Wigner representation and that we
can efficiently draw samples from a Gaussian distribution.  This means that our
results will apply to any measurements that satisfy these properties.
We focus our attention on Gaussian measurements rather than these more
general measurements because of their simplicity and physical relevance.

\section{Simulation Algorithm}\label{sec:sim}

At first glance it may seem that a simulation algorithm for linear
optics may be difficult owing solely to exponential size of the Hilbert
space dimension that is generated by the evolution. We overcome this
problem by exploiting the fact that our Hamiltonians are quadratic
in $p$ and $q$, which implies that the evolution of the Wigner function
follows the Liouville equation as shown in Eq.~\eqref{eq:Wpevol}. Since
the Liouville equation preserves non-negativity and probability mass,
the Wigner function will remain a classical distribution throughout
the evolution. This allows us to model evolution of the Wigner function
using an ensemble of classical trajectories, each of which can be
efficiently simulated. The resulting trajectories can be used to efficiently
draw samples from the final distribution of measurement
outcomes prescribed by the Born rule without needing to know the final quantum state.

It is critical to understand that we are not simulating
the evolution of the quantum state, rather we are simulating measurement
outcomes from a quantum circuit; this is exactly analogous to the
difference between knowing a probability distribution and being able
to sample from it. In particular, the ability to efficiently draw samples does
not imply the ability to efficiently learn the underlying distribution 
because the dimension of the probability
distribution on $n$ modes is exponentially large.


We show in this section that this simulation strategy can be used
to efficiently sample from the output of the following class of quantum
algorithms: 
\begin{algorithmclass}[H]
\begin{enumerate}
\item Apply the linear optical transformation $U_{T,\bm{x}}$. 
\item Perform the separable Gaussian measurement $\measz=\measj{1}\otimes\measj{2}\otimes\dots\otimes\measj{n}$,
where we follow the naming convention of equation \ref{eq:guass_meas_povm}
and the tensor product is understood to mean that the POVM elements
of $\measz=\measj{1}\otimes\cdots\otimes\measj{n}$ are tensor product combinations of
the POVM elements of $\measz$ in the obvious
way.
\item Return the measurement outcome $\bm{k}=(\bm{k}_{1},\bm{k}_{2},\cdots,\bm{k}_{n})\in\mathbb{R}^{2n}$ corresponding to the mean of a Gaussian POVM element. 
\end{enumerate}
\caption{Family of efficiently simulatable quantum algorithms\\ 
\textbf{Input:} Number of modes $n$, an initial $n$ mode quantum state $\rho=\rho_1\otimes\dots\otimes\rho_{n}$
where each $\rho_j$ has positive Wigner representation $W_{\rho_j}(\bm{u})$, a description of a linear optical transformation $U_{T,x}$ which is parameterized by $T\in \mathbb{R}^{2n\times 2n}$ and $\bm{x}\in\mathbb{R}^{2n}$.\\
\textbf{Output:} A string of measurement
outcomes $\bm{k}$ sampled according to the probability density $p(\bm{K}_{\text{quant}}=\bm{k})$
determined by the Born rule.  \label{alg:Simulable-Quantum-Circuit}}
\end{algorithmclass}
{\flushleft Here we conceive of any quantum algorithm in this class
as a way of sampling outcome strings $\bm{k}$ distributed
according to the probability densities given by the Born rule. We
label the corresponding random variable $\bm{K}_{\text{quant}}$.
Here we are \emph{not} simulating the evolution of the Wigner distribution,
which would be equivalent to simulating the full quantum state. Rather,
we show that there is a corresponding classical algorithm that produces
outcome strings $\bm{k}$ with (very nearly) the same distribution those from algorithm class~\ref{alg:Simulable-Quantum-Circuit}.}

Using intuition similar to that in~\cite{Veitch2012Negative}, we note that quantum algorithms in class~\ref{alg:Simulable-Quantum-Circuit}, can be simulated using
the following classical algorithm,
 provided access to classical resources with infinite numerical precision and a blackbox function that can be used to draw samples from $W_{\rho_j}(\bm{u})$ for $j=1,\ldots,n$.  We will later provide an algorithm that does not require infinite precision, but we provide the infinite precision algorithm first because it conveys the necessary intuition without focusing on the technical issues that arise when discretizing the distributions.
\begin{algorithm}[H]
\begin{enumerate}
\item Sample $\bm{u}\in\mathbb{R}^{2n}$ according to the distribution 
$W_{\rho}(\bm{u})=W_{\rho_1}(\bm{u}_{1})\cdots W_{\rho_n}(\bm{u}_{n})$ by drawing a sample independently from each mode using the blackbox function.
\item Apply the affine transformation $\tilde{\bm{u}}=T\bm{u}+\bm{x}$
corresponding to the linear optical transformation $U_{T,\bm{x}}$ to the sampled phase space point $\bm{u}$.
This transformation is an affine mapping due to Louiville's
theorem. 
\item Return the outcome string $\bm{k}=(\bm{k}_{1},\bm{k}_{2},\cdots,\bm{k}_{n})\in\mathbb{R}^{2n}$
from the distribution \\$M_{\measz}(\bm{k}|\tilde{\bm{u}})=M_{\measj{1}}(\bm{k}_{1}|\tilde{\bm{u}}_{1})\cdots M_{\measj{n}}(\bm{k}_{n}|\tilde{\bm{u}}_{n})$,
where $M_{\measj{j}}(\bm{k}_{i}|\tilde{\bm{u}}_{i})$
is given as in equation \ref{eq:gauss_meas_wig_rep}, 
\[
M_{\measj{j}}(\bm{k}_{i}|\tilde{\bm{u}}_{i})=\frac{1}{\mathcal{X}_j}\exp\left(-(\bm{k}_{i}-\tilde{\bm{u}}_{i})^{T}V_{\mathcal{M},j}^{-1}(\bm{k}_{i}-\tilde{\bm{u}}_{i})\right).
\]
\end{enumerate}
\caption{Infinite precision classical simulation algorithm for algorithms in class~\ref{alg:Simulable-Quantum-Circuit}\label{alg:Equivalent-Classical-Circuit}\protect
\protect \\
\textbf{Input:} As algorithms in class~\ref{alg:Simulable-Quantum-Circuit}, except $\rho$ is not provided.\\
\textbf{Output:} A string of measurement
outcomes $\bm{k}$ sampled according to the probability density $p(\bm{K}_{\text{class}}^{\naivename}=\bm{k})$.}  
\end{algorithm}
{\flushleft The intuition behind this class of algorithms is to use
the classical phase space model afforded to us by the non-negative Wigner functions and quadratic evolutions in order to turn
the quantum problem into one that can be efficiently simulated by
a classical computer. In this context we can think of our quantum
system as actually being definitely at some point $\bm{u}\in\mathbb{R}^{2n}$
which is unknown to us. The point then moves under a fully deterministic
classical evolution and measurement on each register amounts to picking
a point $\bm{k}$ from a normal distribution centered at the
location of the system. Each classical algorithm samples outcomes
$\bm{k}$ according the probability density 
\[
p(\bm{K}_{\text{class}}^{\naivename}=\bm{k})=\int_{\bm{u}}W_\rho(\bm{u})M_{\measz}(\bm{k}|\bm{u})d\bm{u},
\]
which agrees with the density given by the Born rule. Thus the outcomes
$\bm{K}_{\text{class}}^{\naivename}$ from the classical
simulator are distributed in exactly the same way as $\bm{K}_{\text{quant}}$, which are outcomes drawn from the actual quantum system.}

Unfortunately, algorithms similar to~\ref{alg:Equivalent-Classical-Circuit} cannot be executed precisely on digital computers and instead would require an analog computer (often referred to as a \emph{real computer}).  If physical, such computers would have unrealistic computational powers such as being able to solve NP--complete problems in polynomial time~\cite{VSD86} and would also violate the holographic principle~\cite{Aar05}.  For these reasons, we need to discretize~\ref{alg:Equivalent-Classical-Circuit} in order to assess the cost of simulating linear optics on realistic classical computers.
The major technical difference between
the continuous variable case and the discrete case considered in \cite{Veitch2012Negative} involves
 showing that finite-precision errors can be made negligible with efficient overhead costs given a set of reasonable assumptions about the input states, dynamics and measurements.

Since infinite precision is requried in the continuous variable setting, in order to specify a quantum state we must make some finite precision truncation. To this end we shall assume access to a family of oracles $\mathcal{W}_{\rho_{j},\eta}(l,m)$ that takes integers $l,m$ and returns a value satisfying  $|\mathcal{W}_{\rho_{j},\eta}(l,m)-W_{\rho_{j}}(\bm{\mu}_{\rho_j}+(l,m)\delta)|<\eta.$ That is, each oracle queries the Wigner function at points on a grid centered at the mean of the distribution. This is a weak assumption as it does not require us to even know the state we are simulating. Using this resource the algorithm can be written as:

\begin{algorithm}[H]
\caption{Finite-precision classical simulation algorithm for quantum algorithms in class~\ref{alg:Simulable-Quantum-Circuit}\label{alg:advanced-Equivalent-Classical-Circuit}\\
\textbf{Input:} As algorithm~\ref{alg:Equivalent-Classical-Circuit}, but also require $\delta$, a discretization length for the input, $|\bm{\epsilon}_2|$, a bound for the numerical error involved in applying the affine transformation, $\Gamma$, a discretization length for the output, $\area$, the area truncated square region of phase space that the simulator considers and $\bm{\mu}_{\rho_j}$, the mean of the Wigner distribution of the quantum state on each mode $j$. We require $\sqrt{\area}$ to be an odd integer multiple of $\delta$ and $\Gamma$ to be an odd integer multiple of $\delta$.\\
\textbf{Output:}  A string of measurement outcomes
$\bm{k}$ sampled according to $\Pr(\bm{K}_{\text{class}}=\bm{k})\equiv\Pr_{\text{sim}}(\bm{k})$.}
\begin{enumerate}
\item For each $j=1,\ldots, n$ execute a through d. (This step approximates sampling a point from phase space.)
\begin{enumerate}
\item For each integer $l,m \in \left[-\frac{\sqrt{\area}}{\delta},\frac{\sqrt{\area}}{\delta}\right]$ set $\Pr_{\text{{\rm sim}},\rho_{j}}(l,m)=\mathcal{W}_{\rho_{j},\eta}(l,m)\cdot\delta^{2}.$ 
\\(The phase space is truncated to a size $\area$ and discretized into boxes of size $\delta$. This step sets a pdf over the centers of the boxes.)

\item For each $(l,m)$ set $\Pr_{\text{{\rm sim}},\rho_{j}}(l,m)=\Pr_{\text{{\rm sim}},\rho_{j}}(l,m)/\sum_{l,m}\Pr_{\text{sim},\rho_{j}}(l,m)$
(This step normalizes the pdf.)

\item Draw a sample $(l,m)$ from the pdf  $\Pr_{\text{{\rm sim}},\rho_{j}}(l,m).$

\item Set $\boldsymbol{u}_{j}=\boldsymbol{\mu}_{j}+(l,m)\delta $
\end{enumerate}

\item Set $\tilde{\bm{u}}=T\bm{u}+\bm{x}$ using enough digits of precision such that the numerical error is at most $|\bm{\epsilon}_{2}|$, where $\bm{u}\equiv\bm{u}_{1}\oplus\dots\oplus\bm{u}_{n}$ and $\tilde{\bm{u}}\equiv\tilde{\bm{u}}_{1}\oplus\dots\oplus\tilde{\bm{u}}_{n}$.\\ (This step corresponds to updating the sampled state according to the linear optical transformation.)

\item  For each $j=1,\ldots, n$ execute a through e. (This step is to simulate drawing a measurement outcome from the Gaussian measurement distribution centered at $\tilde{\bm{u}}$.)

\begin{enumerate}

\item  For each integer $l,m \in \left[-\frac{\sqrt{\area}}{\delta},\frac{\sqrt{\area}}{\delta}\right]$ set  $\Pr_{\text{{\rm sim}},\measj{j}}(l,m):=\exp\left(-\delta(l,m)^{T}V_{\mathcal{M},j}^{-1}\delta(l,m))\right))\cdot\delta^{2}.$
\\(The outcome space is truncated to a size $\area$ and discretized into boxes of size $\delta$. This sets a pdf over the centers of the boxes.)

\item  Set $\Pr_{\text{{\rm sim}},\measj{j}}(l,m)=\Pr_{\text{{\rm sim}},\measj{j}}(l,m)/\sum_{l,m}\Pr_{\text{{\rm sim}},\measj{j}}(l,m).$\\
(This step normalizes the pdf.)

\item Draw a sample $(l,m)$ from the pdf  $\Pr_{\text{{\rm sim}},\measj{j}}(l,m)$.

\item Find integers $(r,s)$ such that $|\delta(l,m)-\Gamma(r,s))|_{\infty}\le\Gamma/2$.\\
(This just amounts to rebinning the outcome distribution into hypercubes of sidelength $\Gamma$;
this introduces no errors but does require $\delta\le\Gamma$.) 

\item Set measurement outcome $\bm{k}_{j}=\tilde{\bm{u}}_j+\Gamma(r,s)$.

\end{enumerate}

\item Return measurement outcome $\bm{k}\equiv\bm{k}_{1}\oplus\dots\oplus\bm{k}_{n}$.

\end{enumerate}
\end{algorithm}


Our simulation protocol can necessarily only sample from a discrete
distribution so we must introduce some notion of how a discrete distribution
can be close to the continuous probability density given by the Born
rule. The most natural way to do this is to discretize the outcome
distribution into boxes of side length $\Gamma$ according to, 
\begin{definition}
Let $\mathcal{N}_{\Gamma}(\bm{k})\subset\mathbb{R}^{2n}$
be a hypercube in outcome space with side length $\Gamma$ centered
at the point $\bm{k}$. We define the $\Gamma$ discretization
of the quantum outcome distribution to be $\Pr(\bm{K}_{\text{quant,\ensuremath{\Gamma}}}=\bm{k})\equiv\Prquant(\bm{k})\equiv\int_{\mathcal{N}_{\Gamma}(\bm{k})}p(\bm{K}_{\text{quant}}=\tilde{\bm{k}})d\tilde{\bm{k}}$. 

We can now fix a $\Gamma$ according to our operational requirements
for the simulation and ask how well a simulation protocol samples
from this distribution. Notice this is an \emph{unavoidable} consequence
of trying to approximate a continuous quantity with a discrete system.
With this in hand we can give a precise classical simulation protocol
by discretizing our naive algorithm, which results in the family of
protocols described in algorithm \ref{alg:advanced-Equivalent-Classical-Circuit}.
\end{definition}

It now easy to see that both the cost of the simulation and the error
in our sampling will be a function of the discretization parameters
$\delta(n,\epsilon)$ and $|\arearegion(n,\epsilon)|$. If we can pick these
parameters such that for fixed error our simulation scheme scales
as ${\rm poly}(n)$ then the simulation is efficient. Concretely,

\begin{definition} Let $\Prsim(\bm{k})$ and $\Prquant(\bm{k})$ be the simulated and actual probabilities of obtaining a measured value that is inside a hypercube of volume $\Gamma^{2n}$ centered at a point in phase space $\bm{k}\in \mathbb{R}^{2n}$ for the separable Gaussian measurement $\measz$.  A simulation algorithm is efficient if for inputs $n,\Gamma$ and $\epsilon$ there exists a choice of $\area,\delta$
such that:
\begin{enumerate}
\item The 1-norm distance between the $\Gamma$-discretized quantum distribution
and the simulator distribution is at most $\epsilon$, 
\[
|\Prquant(\bm{k})-\Prsim(\bm{k})|_{1}\le\epsilon
\]
 where we take $\Prsim(\bm{k})=0$
whenever $\bm{k}$ is outside the domain of $\Prsim(\bm{k})$ defined by algorithm \ref{alg:advanced-Equivalent-Classical-Circuit}. 
\item The computational complexity of the simulation scales as ${\rm poly}(n/\epsilon)$. 
\end{enumerate}
\end{definition} 

We now can show that the simulation of sufficiently smooth separable positive Wigner functions under linear optical operations and Gaussian measurements
is efficient.  This result is formally stated in the following theorem and proof is given in appendix~\ref{appendix:thmproof}.

\begin{theorem}\label{thm:mainthm}   The output of algorithm~\ref{alg:advanced-Equivalent-Classical-Circuit} satisfies $|\Prquant(\bm{k})-\Prsim(\bm{k})|_{1}\le\epsilon$ for input\\ $(\{\bm{\mu}_{\rho_j}\},\{V_{\rho_j}\},\{V_{\mathcal{M},j}\},T,x,\delta,|\bm{\epsilon}_2|,\Gamma,\area)$ if
\begin{enumerate}
\item $n$, $\max_j\{|\bm{\mu}_{\rho_j}|,\|V_{\rho_j},\|V_{\mathcal{M},j}\|\},\|T\|$ and $\|x\|$ are bounded,
\item There exist finite $\beta$, $\Lambda$ such that $|\nabla W_{\rho}(\bm{u})|\le n\beta/\area^{n}$ and $\left|\nabla_{\bm{k}}M_{\measz}(\bm{k}|{\bm{u}})\right|\le n\Lambda/\area^{n}$\label{item:betagammabd}
for $\bm{k}\in\mathbb{R}^{2n},\ \bm{u}\in\mathbb{R}^{2n}$,
\item $\delta\le\min\left\{\frac{\epsilon}{16\left[\left(1+\|T\|\right)\Lambda+\beta\right]n\sqrt{2n}},\Gamma\right\}$,\label{item:deltabound}
\item $\epsilon< 1$, $|\bm{\epsilon}_2|\le \|T\|\delta\sqrt{\frac{n}{2}}$,
\item $\area\ge 16n\max_{i,j}\left(\left[V_{\rho_i}\right]_{11}+\left[V_{\rho_i}\right]_{22}+\left[V_{\mathcal{M},j}\right]_{11}+\left[V_{\mathcal{M},j}\right]_{22}\right)/\epsilon$,\label{item:areabound}
\item The finite precision error from each oracle $\mathcal{W}_{\rho_j,\eta}$ satisfies $\eta\le\frac{\epsilon}{8n\area^{n}}$.
\end{enumerate}
Furthermore, if we assign unit cost to evaluations of $\mathcal{W}_{\rho_j,\eta}$ and Gaussian functions unit cost then the computational complexity of the algorithm is
$O\left(n^{5}\left(\max_i\|V_{\rho_i}\|+\max_j\|V_{\mathcal{M},j}\|\right)\left[\Lambda^2\|T\|^2+\beta^2\right]/\epsilon^3\right),$ which implies efficiency.
\label{thm:main} \end{theorem} 

The key insight of this theorem is that the assumption of Gaussian preparations made in the continuous variable Gottesman--Knill theorem can be relaxed~\cite{Bartlett2002Efficient,Bartlett2002Efficient2}, and that a much wider class of quantum dynamics can be efficiently simulated than previously thought. Indeed, although we stated the algorithm only for product state inputs and product measurements we can now see that this restriction was unnecessary. In fact, our simulation scheme works for any positively represented input and measurement as long as it is possible to efficiently sample from the corresponding distribution. The product assumption is a sufficient but not necessary condition for this efficient sampling. We also note that the algorithm requires us to know the mean and covariance matrices of the distributions, which might be hard to compute analytically. However, since we already require efficient sampling we may appeal to Monte Carlo estimation protocols to determine these quantities within acceptable error tolerances.
  This extension of the continuous variable Gottesman-Knill theorem places much stronger limitations on the input states that
can be used for continuous variable quantum computation and underscores the significance of negativity in the Wigner function as a resource for quantum computation (in analogy to recent results for discrete systems).  In particular, we will show that Theorem~\ref{thm:mainthm} places a minimum efficiency for a class of photonic thermal states beyond which the states cannot be used as a resource for linear optical quantum computation with Gaussian measurements.

\section{Efficient Simulation of Single Photon--Added--Thermal States}\label{sec:SPATS}

The debate over the {}``correct'' definition of classicality for
quantum states of light has been a long and, at times, fierce one.
One of the most common notions of classicality is whether a state
can be represented as a convex combination of Gaussian states. Here
we have shown that, in the context of computational power, such a condition
is superseded by the condition of positivity of the Wigner function.
In this section we give a concrete example of an interesting class of
states which are not Gaussian but which have positive Wigner representation and thereby admit an efficient classical simulation.  

We consider the experimentally accessible class of states
called single-photon-added thermal states (SPATS) \cite{Agarwal1992Nonclassical,Zavatta2004QuantumtoClassical,Parigi2007Probing}:
\[
\rho(\barn)=\frac{1}{\barn(\barn+1)}\sum_{n=0}^{\infty}\left(\frac{\barn}{\barn+1}\right)^{n}n\op{n}{n},
\]
 where $\barn$ is the mean photon number---given by, in terms of
temperature $T$, the Planck distribution $\barn=1/(\exp(1/T)-1)$.
It is known that all states in this class are outside the convex hull
of Gaussian states and have negative Wigner function for finite temperatures.

Under experimentally realistic conditions we must consider states
subjected to losses.  In general, losses can be modeled as an interaction with a vacuum
mode at a beam-splitter with transmittance $\eta$, also called the
\emph{quantum efficiency}. The loss rate is then $1-\eta$.   
The Wigner function of the SPATS after this channel, which we call limited-efficiency SPATS (or LESPATS for short)  is \cite{Zavatta2007Experimental}
\[
W_{\rho(\barn,\eta)}(q,p)=\frac{2}{\pi}\frac{1+2\eta[\barn+2(\barn+1)(q^{2}+p^{2})-2\barn\eta-1]}{(1+2\barn\eta)^{3}}\exp\left(-\frac{2(q^{2}+p^{2})}{1+2\barn\eta}\right).
\]
Note that the most negative value of the LESPATS is at $(q,p)=(0,0)$ for
all $\eta$ and $\barn$. Thus, we consider the quantity 
\[
W_{\rho(\barn,\eta)}(0,0)=\frac{2}{\pi}\frac{1+2\eta(\barn-2\barn\eta-1)}{(1+2\barn\eta)^{3}}.
\]
 By inspection, we can see that for efficiencies of $\eta\leq0.5$, the
Wigner function of the LESPATS is positive $W_{\rho(\barn,\eta)}(q,p)>0$.
Thus, efficiencies of $\eta>0.5$ are necessary for quantum computational speed-up
with SPATS.

Note however that for $\eta\leq0.5$, the LESPATS are not inside the
convex hull of Gaussian states. To see this most clearly, we require
a different phase space distribution. The Glauber $P$-function (see  for example
\cite{Schleich2001Quantum}) is defined as 
\[
\rho=\iint P_{\rho}(q,p)\op\alpha\alpha dqdp,
\]
 where $\ket{\alpha}$ are the \emph{coherent states}, which are vacuum
states that have been displaced in phase space (symmetric Gaussian
states). Note that if $P_{\rho}$ is a probability distribution then
$\rho$ is in the convex hull of coherent states. The $P$-function
of the LESPATS is \cite{Agarwal1992Nonclassical} 
\[
P_{\rho(\barn,\eta)}(q,p)=\frac{1}{\pi\barn^{3}\eta}\left[(\barn+1)\frac{q^{2}+p^{2}}{\eta}-\barn\right]\exp\left(-\frac{q^{2}+p^{2}}{\barn\eta}\right).
\]
 Note that this function is negative for all allowed values of $\eta$
and $\barn$. Thus, the LESPATS are always outsides the convex hull
of Gaussian states but are bound universal states \cite{Veitch2012Negative} for $\eta\leq0.5$.

\begin{figure}[t]
\centering{}{\includegraphics[width=0.9\linewidth]{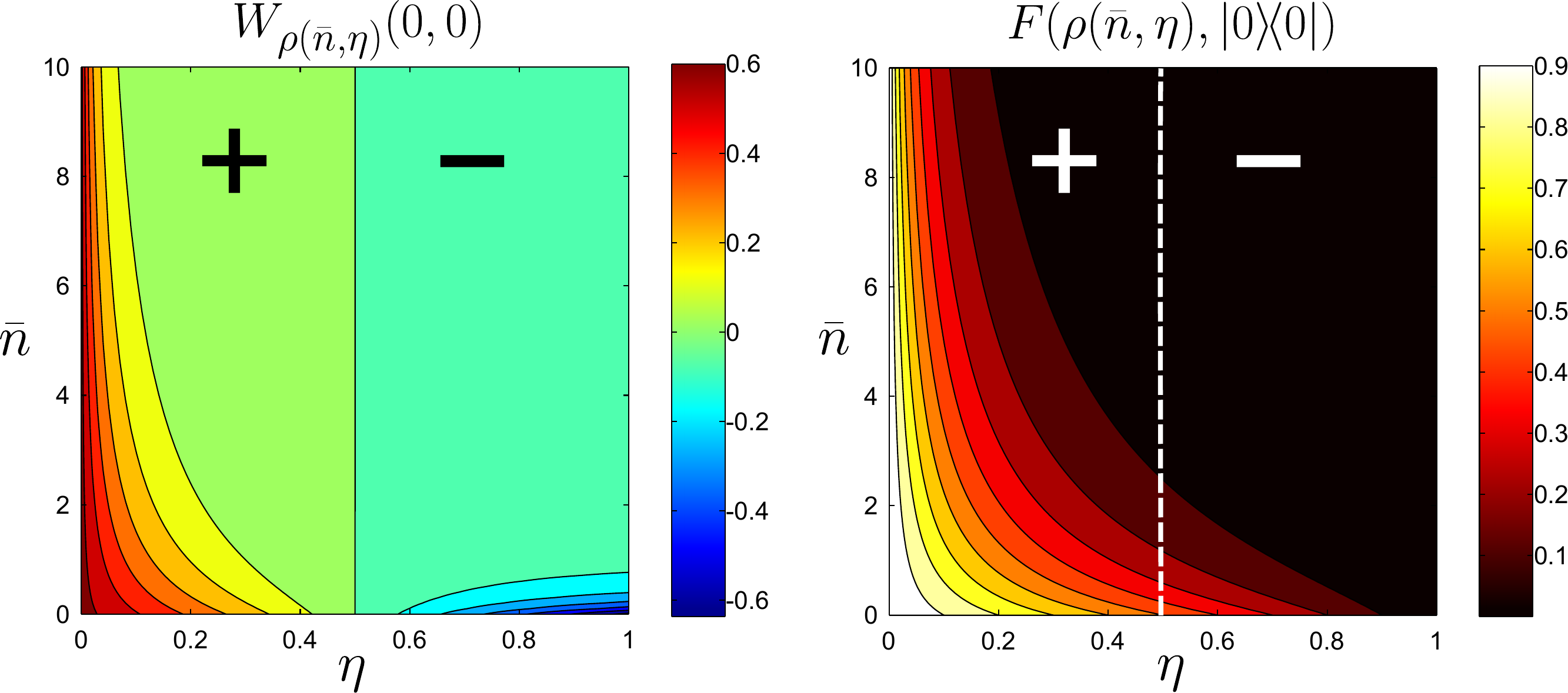}}
\caption{\label{fig:SPATS} Negativity of Wigner function on the left and fidelity to vacuum (``distance'' to convex hull of Gaussians) on the right for varying $\bar{n}$ and $\eta$ (note: $\eta=1$ corresponds to no losses).  In both figures ``+'' indicates the region of non-negative states and ``-'' indicates the region of states with negative Wigner function.  The region of non-negative states ($\eta\leq0.5$) is the region of bound universal states.  This is clear as the Wigner function is positive yet the states still lie
outside the convex hull of coherent states since the $P$-function is always negative.  Notice that the fidelity distance to the convex hull (as measured by the fidelity to the nearest state, $\ket{0}$) is significantly less than 1, suggesting that the region of bound states is quite large.}
\end{figure}

To illustrate this, we compare the negativity of the Wigner function
with the distance to the convex hull of Gaussian states. The distance
we use is based on fidelity, which  can be computed using phase space distributions
as
\[
F(\rho(\barn,\eta),\op00)=\Tr[\rho(\barn,\eta)\op00]=\pi\iint P_{\rho(\barn,\eta)}(q,p)Q_{\op00}(q,p)dqdp,
\]
 where the $Q$-function 
\[
Q_{\op00}(q,p)=\frac{1}{\pi}\exp(-(q^{2}+p^{2}))
\]
 is \emph{dual} to the $P$-function%
\footnote{This duality relationship holds for any phase space representation
\cite{Ferrie2011Quasiprobability}.%
}. Using a standard table of Gaussian integrals we find 
\[
F(\rho(\barn,\eta),\op00)=\frac{1-\eta}{(1+\barn\eta)^{2}}.
\]
 This effect is demonstrated in figure \ref{fig:SPATS}. Note that,
for any state, a quantum efficiency of $\eta\leq 0.5$ is sufficient to ensure
membership of the convex hull of states that have positive Wigner representation. This effect is mirrored
in the discrete case \cite{Veitch2012Negative}, where depolarizing
noise of 50\% is sufficient to ensure membership of the polytope
of states with positive \emph{discrete} Wigner function, when starting from any qudit state.

\section{Conclusion}\label{sec:conclusion}

We have shown that Gaussian quantum computations utilizing separable initial preparations with positive Wigner function are classically efficiently simulable.  Since such states lie outside the convex hull of Gaussian states, we have identified a large class of \emph{bound states}: states that cannot be prepared using Gaussian operations, yet do not permit universal quantum computation.  We illustrated this class using the example of single-photon-added-thermal-states, showing that quantum efficiencies of 50\% are necessary for quantum computation.

Effort has been extended beyond qualitatively defining negativity as quantumness to \emph{quantifying} quantumness via negativity.  In terms of the Wigner function, the \emph{volume} of the negative parts of the represented quantum state has been suggested as the appropriate measure of quantumness \cite{Kenfack2004Negativity}.  The \emph{distance} (in some some preferred norm on the space of Hermitian operators) to the convex subset of positive Wigner functions was suggested to quantify quantumness in reference \cite{Mari2010Directly}.  The \emph{volume} of negativity of the Wigner function (and, in the finite dimensional case, the sum of the negative values) is a ``good'' measure of non-classicality since it is monotonic under Gaussian operators; that is, Gaussian operations cannot increase the volume of the negative regions in phase space.  

Reference \cite{Bartlett2003Requirement} nicely summarized what was
known at the time about continuous variable quantum computation. The
table presented there is reproduced below in table \ref{table:review}
with some more recent results. The field began with Lloyd and Braunstein's
observation that non-linear optical processes are sufficient for universal
continuous variable quantum computation. Later, it was shown for discrete
variable encodings that linear optics is sufficient provided photon
counting measurements are available \cite{Knill2001Scheme,Gottesman2001Encoding}.
The continuous variable analog of the \emph{measurement-based} model
shows that preparation of single photon state preparation is also
sufficient \cite{Menicucci2006Universal}. More recently, the result
of Aaronson and Arkhipov \cite{Aaronson2010Computational} shows that
preparing and measuring single photon states (without post-selection)
is equivalent to a sampling problem that is thought to be hard classically---but
it still manages to (probably) not be universal for quantum computation.

It is possible that the Aaronson and Arkhipov model may be intermediately
between classically efficiently simulatable and universal for quantum
computation. Another suspected model of this type is the {}``one-clean-qubit''
model of Knill and Laflamme \cite{Knill1998Power}. The key point
for this latter model is that uses highly mixed states. Mixed states
have not been given full consideration for continuous variable quantum
computation. Here we have shown, via the Wigner phase space formalism
and independent of purity, negative representation is necessary for
universal quantum computation. Moreover, any computation that uses
states possessing a positive Wigner function is classically efficiently
simulatable. It would be quite interesting if this condition turned
out also to be sufficient as this would provide a sharp boundary between
quantum and classical systems with regard to their computational power.

\begin{acknowledgements}
 We thank Earl Campbell and Christian
Weedbrook for helpful comments. The authors acknowledge financial
support from the Government of Canada through NSERC, CIFAR, USARO-DTO. After completion
of this work, we were made aware of \cite{Mari2012Positive}, who
obtain a similar result for a different (namely, local) set of dynamical
transformations.
\end{acknowledgements}

\appendix
\section{Proof of Theorem~\ref{thm:main}}\label{appendix:thmproof}

\begin{proofof}{Theorem~\ref{thm:main}}
Our goal is to show that with this choice of discretization parameters
$\delta$ and $\area$ the simulation algorithms outlined in algorithm
 \ref{alg:advanced-Equivalent-Classical-Circuit} require $O\!\left({\rm poly}(n/\epsilon)\right)$
resources and result in error at most $\epsilon$. Since we bin the data at the end of the protocol into hypercubes of volume $\Gamma^{2n}\ge \delta^{2n}$ and $\Gamma$ is an integer multiple of $\delta$, no error is introduced by first binning the outcomes into hypercubes of volume $\delta^{2n}$ because every hypercube of volume $\delta^{2n}$ is contained in precisely one hypercube of volume $\Gamma^{2n}$.
 We therefore may take the quantum
distribution to be binned into hypercubes of side length $\delta$
(ie. $\Gamma=\delta$) without loss of generality.

We start by analyzing the error. Following scheme outlined above we
can decompose this into four broad parts: 
\begin{enumerate}
\item The error introduced by the use of finite precision output of $\mathcal{W}_{\rho_j,\eta}$.
\item The error introduced by truncating the sampling distribution over
phase space.
\item The error introduced by discretizing this truncated distribution.
\item The error introduced by truncating the outcome distribution.
\item The error introduced by discretizing the outcome distribution.
\end{enumerate}

Denoting the region in phase space that the initial states are confined to as $\arearegion_{\rho}=\arearegion_{\rho_1}\otimes\cdots\otimes\arearegion_{\rho_n}$ and the region that the observations are confined to as $\arearegion_{\measz}=\arearegion_{\measj{1}}\otimes \cdots \otimes \arearegion_{\measj{n}}$ (where  $|\arearegion_{\rho_i}|=|\arearegion_{\measj{i}}|=\area$),  we can use the triangle inequality to express these errors as:

\begin{align}
|\Prsim(\bm{k})-\Prquant(\bm{k})|_{1}  \le~&\left|\Prsim(\bm{k})-\Prtruncquant(\bm{k})\right|_1\nonumber\\
 & \qquad+\Pr_{\text{quant}}(\bm{u}\notin \arearegion_{\rho})+\text{Pr}_{\text{quant}}(\bm{k}\notin \arearegion_{\measz}),
\label{eq:totalerror}
\end{align}
which says that the total error is at most the distance between the
truncated distributions plus the probability that a point is sampled, or measured, outside the truncated region.  In other words, the 1--norm error introduced by truncating is, even in the most pathological case conceivable, the sum of the probabilities of sampling an initial trajectory outside of $\arearegion_\rho$ and measuring an outcome outside of $\arearegion_{\measj{i}}$.

It then follows that $|\Prsim(\bm{k})-\Prquant(\bm{k})|_{1} \le \epsilon/2$ if
\begin{align}
\Pr_{\text{quant}}(\bm{u}\notin\arearegion_{\rho})+\text{Pr}_{\text{quant}}(\bm{k}\notin\arearegion_{\measz})&\le \epsilon/4,\label{eq:truncerror}\\
\left|\Prsim(\bm{k})-\Prtruncquant(\bm{k})\right|_1&\le \epsilon/4\label{eq:simerror}.
\end{align}
We will first demonstrate that~\eqref{eq:truncerror} is an immediate consequence of assumption \ref{item:areabound}.  We then will show that~\eqref{eq:simerror} is satisfied if assumption~\ref{item:deltabound} holds.

We begin by bounding the truncation error. For the $i^{\rm th}$ mode, 
\begin{align*}
\Pr_{\text{quant}}\left(\bm{u}_{i}\notin\arearegion_{\rho_i}\right) & =  \Pr\left(|q_{i}-\mu_{Q}|>\frac{\sqrt{\area}}{2},|p_{i}-\mu_{P}|>\frac{\sqrt{\area}}{2}\right)\\
 & \le \Pr\left(|q_{i}-\mu_{Q}|>\frac{\sqrt{\area}}{2}\right)+\Pr\left(|p_{i}-\mu_{P}|>\frac{\sqrt{\area}}{2}\right).
\end{align*}
Our upper bounds for both of these probabilities are established using Chebyshev's inequality.  Chebyshev's inequality states that, for a probability distribution $P$ with mean $\mu$ and standard deviation $\sigma$ that $P(|x-\mu|\ge k\sigma)\le (\sigma/k)^2$.  In our case, the two standard deviations are $\sqrt{[V_{\rho_i}]_{11}}$ and $\sqrt{[V_{\rho_i}]_{22}}$ which gives us
\begin{equation}
\Pr_{\text{quant}}\left(\bm{u}_{i}\notin\arearegion_{\rho_i}\right)\le \frac{4([V_{\rho_i}]_{11}+[V_{\rho_i}]_{22})}{\area}.
\end{equation}
It immediately follows from the independence of the $n$ distributions that compose $W_\rho$ that,
\[
\Pr_{\text{quant}}(\bm{u}\notin\arearegion_{\rho})\le4n\max_{i=1,\ldots,n}\left(\frac{\left[V_{\rho_i}\right]_{11}+\left[V_{\rho_i}\right]_{22}}{\area}\right).
\]
By identical reasoning,
\[
\Pr_{\text{quant}}(\bm{k}\notin\arearegion_{\measz})\le4n\max_{j=1,\ldots,n}\left(\frac{[V_{\mathcal{M},j}]_{11}+[V_{\mathcal{M},j}]_{22}}{\area}\right).
\]
Thus we have, 
\[
\text{Pr}_{\text{quant}}(\bm{k}\notin\arearegion_{\rho})+\Pr_{\text{quant}}(\bm{u}\notin\arearegion_{\measz})\le4n\max_{i,j}\left(\frac{\left[V_{\rho_i}\right]_{11}+\left[V_{\rho_i}\right]_{22}+[V_{\mathcal{M},j}]_{11}+[V_{\mathcal{M},j}]_{22}}{\area}\right)
\]
so choosing $\area\ge8n\max_{i,j}\left(\frac{\left[V_{\rho_i}\right]_{11}+\left[V_{\rho_i}\right]_{22}+[V_{\mathcal{M},j}]_{11}+[V_{\mathcal{M},j}]_{22}}{\epsilon/2}\right)$
guarantees, 
\begin{equation}
\Pr_{\text{quant}}(\bm{u}\notin\arearegion_{\rho})+\text{Pr}_{\text{quant}}(\bm{k}\notin\arearegion_{\measz})\le\frac{\epsilon}{4}.\label{eq:truncation_bound}
\end{equation}

We must now bound the discretization error on the truncated distributions.
It is natural to break this error up into three pieces corresponding
to the first three steps of the algorithm. First, there is the error introduced
by discretizing the initial sampling distribution over phase space.
Next there is the numerical error introduced in implementing
the affine transformation $\tilde{\bm{u}}=T\bm{u}+\bm{x}$.
Finally, there is the error introduced by discretizing the outcome
distribution. The remainder of the proof is devoted to bounding these
three errors and thereby bounding the total error using~\eqref{eq:totalerror}.

Step 1 of the simulation algorithm breaks up $\arearegion_{\rho}$
into hypercubes of volume $\delta^{2n}$ and samples from the set
of centers of these hypercubes according to,
\begin{align}
\Prsimsamp(\boldsymbol{u}=\boldsymbol{\mu}+(l_{1},\ldots,l_{n},m_{1},\ldots,m_{n})\delta)&= \Pr_{\text{{\rm sim}},\rho_{1}}(l_{1},m_{1})\cdots\Pr_{\text{{\rm sim}},\rho_{n}}(l_{n},m_{n})\nonumber\\
 &=\mathcal{W}_{\rho_{1},\eta}(l_{1},m_{1})\cdots\mathcal{W}_{\rho_{n},\eta}(l_{n},m_{n})\cdot\delta^{2n}\nonumber\\
&=W_{\rho_{1}}(\boldsymbol{u}_{1})\cdots W_{\rho_{n}}(\boldsymbol{u}_{n})\cdot\delta^{2n}+\bm{\epsilon}_1,
\end{align}
where $\bm{\epsilon}_1$ is the numerical error introduced by using $\eta>0$.
The probability weighting assigned to a hypercube center $\bm{u}$
is only approximately equivalent to the probability mass contained in the hypercube;
we must bound the error introduced by this nonequivalence. Define $\mathcal{N}_{\delta}(\bm{u})\subset\mathbb{R}^{2n}$
to be the hypercube with side length $\delta$ centered at the point
$\bm{u}$. The for a fixed $\bm{u}$ in the domain
of $\Prsimsamp(\bm{u})$ we have that every point $\bm{w}\in \mathcal{N}_\delta(\bm{u})$ is also in the region of truncation $\arearegion$ hence $W_{\rho_j}^{\rm trunc}(\bm{w})=W_{\rho_j}(\bm{w})$ for all such $\bm{w}$.  We then use this simplifying observation, the mean value theorem and the triangle inequality to find,

\begin{align}
\left|\Prsimsamp(\bm{u})-\int_{\mathcal{N}_{\delta}(\bm{u})}W_{\rho}(\bm{v})d\bm{v}\right| & \le  \left(\max_{\bm{v}\in\mathcal{N}_{\delta}(\bm{u})}|\nabla W_{\rho}(\bm{v})|\right)\delta^{2n}\max_{\bm{v}\in\mathcal{N}_{\delta}(\bm{u})}|\bm{v}-\bm{u}|+|\bm{\epsilon}_1|\nonumber\\
 & \le  \frac{\delta^{2n+1}\beta n\sqrt{n/2}}{\area^{n}}+|\bm{\epsilon}_1|.\label{eq:term_1_bound}
\end{align}
Where $\max_{\bm{v}\in\mathcal{N}_{\delta}(\bm{u})}|\bm{v}-\bm{u}|\le\sqrt{2n}\frac{\delta}{2}$
from Pythagoras' theorem and $\left(\max_{\bm{v}\in\mathcal{N}_{\delta}(\bm{u})}|\nabla W_{\rho}(\bm{v})|\right)\le n\beta/\area^n$ by assumption~\ref{item:betagammabd} of Theorem~\ref{thm:mainthm}. 

In the second step of the simulation algorithm we move the sampled
point $\bm{u}\in\mathbb{R}^{2n}$ to the point $\tilde{\bm{u}}=T\bm{u}+\bm{x}+\bm{\epsilon}_{2}$,
simulating the evolution due to the linear optical transformation
$U_{T,\bm{x}}$. The numerical error $\bm{\epsilon}_{2}$ depends
only on numerical precision which can be made exponentially small
using a linear amount of memory. The other source of error in the simulation of the dynamics is due to error in the initial conditions caused by sampling the point $\bm{u}\in \mathbb{R}^{2n}$ as opposed to the point $\bm{v}\in \mathbb{R}^{2n}$ that would have been sampled if $\delta=0$.  The triangle inequality then implies that
\begin{equation}
|\tilde{\bm{u}}-\tilde{\bm{v}}|:=\bm|\left(T\bm{u}+\bm{x}+|\bm{\epsilon}_{2}|\right)-\left(T\bm{v}+\bm{x}\right)|\le|\bm{\epsilon}_2|+\|T\|\max_{\bm{v}\in\mathcal{N}_{\delta}(\bm{u})}|\bm{v}-\bm{u}| = |\bm{\epsilon}_2|+\|T\|\delta\sqrt{\frac{n}{2}}.\label{eq:evol_error_bound}
\end{equation}
Since $\bm{\epsilon}_2$ can be made exponentially small using a polynomial number of computational steps, we can choose $|\bm{\epsilon}_2|\le \|T\|\delta\sqrt{\frac{n}{2}}$ without affecting the efficiency of the algorithm.  Therefore, by making such a choice, the total error in the simulated dynamics is at most 
\begin{equation}
|\tilde{\bm{u}}-\tilde{\bm{v}}|\le \|T\|\delta\sqrt{{2n}}.
\end{equation}

The final step of the simulation algorithm samples a measurement outcome
on each \registerword. To do so we break up the truncated outcome
space $\arearegion_{\measz}$ into hypercubes of
volume $\delta^{2n}$ and sample from the set
of centers of these hypercubes according to,
\begin{align}
\Prsim(\bm{k}=\tilde{\boldsymbol{\mu}}+(l_{1},\ldots,l_{n},m_{1},\ldots,m_{n})\delta|\tilde{\bm{u}})  &=  \Pr_{\text{{\rm sim}},\measj{1}}(l_{1},m_{1})\cdots\Pr_{\text{{\rm sim}},\measj{n}}(l_{n},m_{n})\nonumber\\ 
&=\frac{1}{\zeta}\exp\left(-\sum_{i=1}^{n}\delta(l_{i},m_{i})^{T}V_{\mathcal{M},i}^{-1}\delta(l_{i},m_{i}))\right)\cdot\delta^{2n}\nonumber\\
&=\frac{1}{\zeta}\exp\left(-\sum_{i=1}^{n}(\bm{k}_{i}-\tilde{\bm{u}}_{i})^{T}V_{\measj{i}}^{-1}(\bm{k}_{i}-\tilde{\bm{u}}_{i})\right)\cdot\delta^{2n},
\end{align}
where $\zeta$ is a normalizing constant and $\bm{k}_i$ are components of $\bm{k}$.
 As in step one we
may use the mean value theorem to bound the error introduced by sampling
this way rather than according to the true probability mass over each
hypercube. Using $\mathcal{N}_{\delta}(\bm{k})\subset\mathbb{R}^{2n}$
to be the hypercube with side length $\delta$ centered at the point
$\bm{k}$ we find from the mean value theorem and the assumptions of theorem~\ref{thm:mainthm} that the error in the probability enclosed in a single hypercube centered
at $\bm{k}$ is at most:

\begin{eqnarray}
\left|\Pr_{\text{{\rm sim}},\measz}(\bm{k}|\tilde{\bm{u}})-\int_{\mathcal{N}_{\delta}(\bm{k})}M_{\measz}(\boldsymbol{\kappa}|\tilde{\bm{u}})d\boldsymbol{\kappa}\right| & \le & \max_{\boldsymbol{\kappa}\in\arearegion_{\measz}} \left|\nabla_{\boldsymbol{\kappa}} M_{\measz}(\boldsymbol{\kappa}|\tilde{\bm{u}})\delta^{2n}\right|\max_{\boldsymbol{\kappa}\in\mathcal{N}_{\delta}(k)}|\boldsymbol{\kappa}-\bm{k}|\nonumber\\
 & \le & \frac{\Lambda\delta^{2n+1}n\sqrt{2n}}{\area^{n}}\label{eq:meas_error_bound}.
\end{eqnarray}

We complete the error analysis by bounding the distance between the simulator
distribution and truncated, discretized quantum distribution. Our
aim is to rewrite this error in terms of quantities we have already
bounded. Let $\mathcal{D}_{\bm{u}}$ be the domain of $\Prsim(\bm{u})$;
ie. the discrete set of points in $\arearegion_\rho$ that  can be sampled
by the simulator. The difference between the probabilities of a fixed outcome $\bm{k}$ occurring for both the simulation and the truncated quantum distribution

\begin{align}
&\left|\Prsim(\bm{k})-\Prtruncquant(\bm{k})\right|\nonumber\\
  &\qquad\qquad=\left|\sum_{\bm{u}\in\mathcal{D}_{\bm{u}}}\Prsim(\bm{k}|\tilde{\bm{u}}+\bm{\epsilon}_{2})\Prsimsamp(\bm{u})-\int_{\mathcal{N}_{\delta}(\bm{k})}\int_{\arearegion_{\rho}}M_{\measj{i}}^{\text{trunc}}(\boldsymbol{\kappa}|\tilde{\bm{v}})W_{\rho}(\bm{v})\mathrm{d}\bm{v}\mathrm{d}\boldsymbol{\kappa}\right|\nonumber \\
 & \qquad\qquad=\left|\sum_{\bm{u}\in\mathcal{D}_{\bm{u}}}\Prsim(\bm{k}|\tilde{\bm{u}}+\bm{\epsilon}_{2})\Prsimsamp(\bm{u})-\sum_{\bm{u}\in\mathcal{D}_{\bm{u}}}\int_{\mathcal{N}_{\delta}(\bm{k})}\int_{\mathcal{N}_{\delta}(\bm{u})}M_{\measj{i}}^{\text{trunc}}(\boldsymbol{\kappa}|\tilde{\bm{v}})W_{\rho}
(\bm{v})\mathrm{d}\bm{v}\mathrm{d}\boldsymbol{\kappa}\right|\nonumber\\
 & \qquad\qquad\le\sum_{\bm{u}\in\mathcal{D}_{\bm{u}}}\Prsim(\bm{k}|\tilde{\bm{u}}+\bm{\epsilon}_{2})\left|\Prsimsamp(\bm{u})-\int_{\mathcal{N}_{\delta}(\bm{u})}W_{\rho}(\bm{v})\mathrm{d}\bm{v}\right|\nonumber \\
 & \qquad\qquad\qquad+\sum_{\bm{u}\in\mathcal{D}_{\bm{u}}}\int_{\mathcal{N}_{\delta}(\bm{u})}\left|\Prsim(\bm{k}|\tilde{\bm{u}}+\bm{\epsilon}_{2})-\int_{\mathcal{N}_{\delta}(\bm{k})}M_{\measj{i}}^{\text{trunc}}(\boldsymbol{\kappa}|\tilde{\bm{v}})\mathrm{d}\boldsymbol{\kappa}\right|W_{\rho}(\bm{v})\mathrm{d}\bm{v}.\label{eq:samperr}
\end{align}
 The final inequality is found by adding and subtracting $\sum_{\bm{u}\in\mathcal{D}_{\bm{u}}}\Prsim(\bm{k}|\tilde{\bm{u}}+\bm{\epsilon}_{2})\int_{\mathcal{N}_{\delta}(\bm{u})}W_{\rho}(\bm{v})\mathrm{d}\bm{v}$
and applying the triangle inequality.

We need one more intermediary bound before arriving at the final result.
Let $\bm{v}\in\mathcal{N}_{\delta}(\bm{u})$. It follows from
the mean value theorem and $M_{\measz}^{\text{trunc}}=M_{\measz}$ throughout the domain of integration that
\begin{align}
&\left|\int_{\mathcal{N}_{\delta}(\bm{k})}M_{\measz}^{\text{trunc}}(\boldsymbol{\kappa}|\tilde{\bm{u}}+\bm{\epsilon}_{2})d\boldsymbol{\kappa}-\int_{\mathcal{N}_{\delta}(\bm{k})}M_{\measz}^{\text{trunc}}(\boldsymbol{\kappa}|\tilde{\bm{v}})d\boldsymbol{\kappa}\right| \nonumber\\
&\qquad\qquad=\left|\int_{\mathcal{N}_{\delta}(\bm{k})}M_{\measz}(\boldsymbol{\kappa}|\tilde{\bm{u}}+\bm{\epsilon}_{2})d\boldsymbol{\kappa}-\int_{\mathcal{N}_{\delta}(\bm{k})}M_{\measz}(\boldsymbol{\kappa}|\tilde{\bm{v}})d\boldsymbol{\kappa}\right| \nonumber\\
&\qquad\qquad \le \int_{\mathcal{N}_{\delta}(\bm{k})}\max_{\bm{x}\in\arearegion_{\rho}}\left|\nabla_{\bm{u}}M_{\measz}^{\text{trunc}}(\boldsymbol{\kappa}|\bm{u})|_{\bm{u}=\bm{x}}\right|\cdot\left|T(\bm{u}-\bm{v})+\bm{\epsilon}_{2}\right|d\boldsymbol{\kappa}\nonumber\\
 &\qquad\qquad \le  \int_{\mathcal{N}_{\delta}(\bm{k})}\frac{n\Lambda}{\area^{n}}\cdot|T(\bm{u}-\bm{v})+\bm{\epsilon}_{2}|d\boldsymbol{\kappa}\nonumber\\
 &\qquad\qquad \le  \delta^{2n}\frac{n\Lambda}{\area^{n}}\left(\|T\|\delta\sqrt{\frac{n}{2}}+\|T\|\delta\sqrt{\frac{n}{2}}\right),
\end{align}
where we have used the assumption that $|\bm{\epsilon}_2|\le \|T\|\delta\sqrt{n/2}$
and identical reasoning to bound $\left|T(\bm{u}-\bm{v})\right|$.
In conjunction with \eqref{eq:meas_error_bound} this implies:
\begin{equation}
\left|\Prsim(\bm{k}|\tilde{\bm{u}}+\bm{\epsilon}_{2})-\int_{\mathcal{N}_{\delta}(\bm{k})}M_{\measz}^{\text{trunc}}(\boldsymbol{\kappa}|\tilde{\bm{v}})\mathrm{d}\boldsymbol{\kappa}\right|\le\left(1+\|T\|\right)\frac{n\Lambda\sqrt{2n}\delta^{2n+1}}{\area^{n}}+|\bm{\epsilon}_1|.\label{eq:term_2_bound}
\end{equation}

Since there are numerical errors in the computation of the probability distribution for measurement outcomes, it is conceivable that the algorithm assigns more than unit probability to outcomes in the region $\mathcal{A}$ (although this is unlikely for small values of $\epsilon$).  Although we can claim that $\sum_{\bm{u}\in\mathcal{D}_{\bm{u}}}\int_{\mathcal{N}_{\delta}(\bm{u})}W_{\rho}(\bm{v})\mathrm{d}\bm{v}\le 1$  we can only claim that $\sum_{\bm{u}\in\mathcal{D}_{\bm{u}}}\Prsim(\bm{k}|\tilde{\bm{u}}+\bm{\epsilon}_{2})\le1+\epsilon \le 2$
under the assumption that $\epsilon< 1$.  Using these facts, we
substitute \eqref{eq:term_1_bound} and \eqref{eq:term_2_bound}
into \eqref{eq:samperr} to find:
\[
\left|\Prsim(\bm{k})-\Prtruncquant(\bm{k})\right|\le\left[\left(1+\|T\|\right)\Lambda+\beta\right]\frac{n\sqrt{2n}\delta^{2n+1}}{\area^{n}}+|\bm{\epsilon}_1|.
\]
Since we have assumed that the outcomes are discretized into hypercubes of volume $\delta^{2n}$, the discretization produced $\frac{\area^{n}}{\delta^{2n}}$ hypercubes. This implies that the $1$-norm distance between the the two distributions is:
\[
\sum_{\bm{k}}\left|\Prsim(\bm{k})-\Prtruncquant(\bm{k})\right|\le\left[\left(1+\|T\|\right)\Lambda+\beta\right]n\sqrt{2n}\delta+|\bm{\epsilon}_1|\area^n/\delta^{2n}.
\]
Since there are $n$ modes and numerical error $\eta$, it is straightforward to see that $|\bm{\epsilon}_1|\le n\eta\delta^{2n} $;
 therefore choosing $$\delta\le\frac{\epsilon}{8\left[\left(1+\|T\|\right)\Lambda+\beta\right]n\sqrt{2n}},$$ and $$\eta\le \frac{\epsilon}{8n\area^{n}},$$
implies
\begin{equation}
|\Prsim(\bm{k})-\Prtruncquant(\bm{k})|_{1}\le\frac{\epsilon}{4}\label{eq:discretization_bound}.
\end{equation}
This sets the discretization error to be $\epsilon/4$.  We found previously that the truncation error is at most $\epsilon/4$ under assumption~\ref{item:areabound} of theorem~\ref{thm:main}.  We then combine \eqref{eq:totalerror}, \eqref{eq:truncation_bound}
and \eqref{eq:discretization_bound} and find:
\[
|\Prsim(\bm{k})-\Prquant(\bm{k})|_{1}\le\epsilon/2,
\]
if $\Gamma=\delta$.
This also implies that the result holds for $\Gamma\ge\delta$ given that $\Gamma$ is an integer multiple of $\delta$ (which ensures that the hypercube of size $\delta^{2n}$ that the measurement is assigned to is inside the hypercube of size $\Gamma^{2n}$ that it should be assigned to given $\Gamma$--discretization); therefore, it generally holds if
\begin{equation}
\delta\le\min\left\{\frac{\epsilon}{8\left[\left(1+\|T\|\right)\Lambda+\beta\right]n\sqrt{2n}},\Gamma\right\},\label{eq:deltabd}
\end{equation}

A final issue remains: we did not consider errors that are introduced in the sampling steps in the algorithm that arise because the simulated probability distributions are not normalized.  We will show that, from the assumption that $\epsilon< 1$, it suffices to divide $\epsilon$ by $2$ in all previous calculations.  To see this, let $\bm{y}$ be a discrete probability distribution and let $\bm{x}$ be a vector such that $|\bm{y}-\bm{x}|_1\le \epsilon'$ for some $1>\epsilon'\ge 0$.  It then follows from the triangle inequality that $1+\epsilon'\ge |\bm{x}|_1\ge 1-\epsilon'$.  Thus
\begin{align}
\big|\bm{y}-\bm{x}/|\bm{x}|_1\big|_1&=\big|\bm{y}|\bm{x}|_1-\bm{x}\big|_1/|\bm{x}|_1\nonumber\\
&=\big|\bm{y}|\bm{x}|_1-\bm{x}|\bm{x}|_1+\bm{x}|\bm{x}|_1-\bm{x}\big|_1/|\bm{x}|_1\nonumber\\
&\le|\bm{y} -\bm{x}|_1+ ||\bm{x}|_1-1|\le 2\epsilon',
\end{align}
under the assumption that $\epsilon'<1$.  Therefore, by combining these results with~\eqref{eq:deltabd}  we see that the difference between the (now normalized) simulated probabilities and the quantum predictions is at most $\epsilon$ if
\begin{equation}
\delta\le\min\left\{\frac{\epsilon}{16\left[\left(1+\|T\|\right)\Lambda+\beta\right]n\sqrt{2n}},\Gamma\right\},\label{eq:deltabd2}
\end{equation}
and
\begin{equation}
\area\ge 16n\max_{i,j}\left(\left[V_{\rho_i}\right]_{11}+\left[V_{\rho_i}\right]_{22}+\left[V_{\mathcal{M},j}\right]_{11}+\left[V_{\mathcal{M},j}\right]_{22}\right)/\epsilon,\end{equation}
as claimed by theorem~\ref{thm:mainthm}.

We complete the proof by showing that algorithm~\ref{alg:advanced-Equivalent-Classical-Circuit}
is computationally efficient with this choice of $\area$ and $\delta$.
Again we will analyze the simulation algorithms step by step. In the
first step we sample a point $\bm{u}_{i}$ on the phase space
of each register from the distribution $\Pr_{\text{sim},\rho}(\bm{u}_{i})$.
This distribution has support on $\frac{\area}{\delta^{2}}$ squares, and thus if we take $\area$ and $\delta$ to be proportional to their respective lower and upper bounds then the number of times that $\mathcal{W}_{\rho_{j},\eta}$ must be evaluated scales as $\Theta\left(n^{4}\left(\|V_\rho\|+\|V_{\mathcal{M}}\|\right)\left[\Lambda^2\|T\|^2+\beta^2\right]/\epsilon^3\right)$ (here $\Theta(\cdot)$ is Bachmann--Landau notation meaning asymptotically bounded above and below by a constant multiplied by $(\cdot)$).  If we ascribe unit computational cost to every such access, then the computational complexity of this step is proportional to the number of times that $\mathcal{W}_{\rho_{j},\eta}$ is queried.  This task must be repeated $n$ times, and hence the total computational complexity of this step is.
$$\Theta\left(n^{5}\left(\|V_\rho\|+\|V_{\mathcal{M}}\|\right)\left[\Lambda^2\|T\|^2+\beta^2\right]/\epsilon^3\right).$$

In step two we apply the affine transformation $\bm{u}\rightarrow T\bm{u}+\bm{x}$,
whose cost is dominated by the cost of performing a matrix multiplication using $O(\log(\|T\|\delta\sqrt{n}))$ bits of precision.  Since the matrix multiplication requires a number of arithmetic operations that scales quadratically with the matrix dimension and addition and multiplication scale at most quadratically with the number of bits of precision, the total cost of this step is $O(n^{2}\log^2(\|T\|\delta\sqrt{n}))$, which is subdominant to the cost of the previous step and therefore does not affect the scaling.

In step three, we are confronted with the task of measuring the resultant trajectory using a separable Gaussian measurement.  
There is a strong duality between drawing a sample trajectory from the initial separable Wigner function and drawing a measurement outcome for the separable Gaussian measurement of the final trajectory.  In particular, we discretize the space surrounding the outcome space of each of the $n$ separable measurements into $\area/\delta^2$ points.  The approximate calculation of the measurement probability requires that we perform a number of operations that are proportional to the number of points.  Therefore, identically to step one, the computational cost is $$\Theta\left(n^{5}\left(\|V_\rho\|+\|V_{\mathcal{M}}\|\right)\left[\Lambda^2\|T\|^2+\beta^2\right]/\epsilon^3\right),$$
which verifies the computational complexity claimed by theorem~\ref{thm:mainthm} and shows that
under the assumptions of the theorem all three steps are computationally
efficient, and hence the algorithm is efficient as well.

\end{proofof}

\begin{center}
\begin{table}[b]
\begin{tabular}{p{3.75cm}|p{3.75cm}|p{3.75cm}|p{3.75cm}}
\toprule \textbf{Preparations}  & \textbf{Gates}  & \textbf{Measurement}  & \textbf{Efficiently simulatable classically} \tabularnewline
\toprule Vacua  & Linear optics  & Gaussian  & {\color{green}\cmark} \cite{Bartlett2002Efficient,Bartlett2002Efficient2}\tabularnewline
\hline 
Vacua  & Non-linear optics  & Gaussian  & {\color{red}\xmark} \cite{Lloyd1999Quantum}\tabularnewline
\hline 
Single photons  & Linear optics (no squeezing)  & Photon counting (with post-selection)  & {\color{red}\xmark} \cite{Knill2001Scheme}\tabularnewline
\hline 
Vacua  & Linear optics  & Gaussian and Photon counting (with post-selection)  & {\color{red}\xmark} \cite{Gottesman2001Encoding}\tabularnewline
\hline 
Single photons  & Linear optics  & Gaussian  & {\color{red}\xmark} \cite{Gu2009Quantum}\tabularnewline
\hline 
Single photons  & Linear optics (no squeezing)  & Photon counting  & {\color{red}\xmark} \cite{Aaronson2010Computational}\tabularnewline
\hline 
Product Positive Wigner functions  & Linear optics  & Product Gaussian  & {\color{green}\cmark} (this work)\tabularnewline
\toprule  &  &  & \tabularnewline
\end{tabular}\caption{\label{table:review} An extension of the table appearing in \cite{Bartlett2003Requirement}.}
\end{table}

\par\end{center}

\bibliographystyle{apsrev}
\bibliography{infinited2}

\end{document}